\documentclass[twocolumn,10pt,aps,pra,showpacs,superscriptaddress,nobalancelastpage,longbibliography,nofootinbib,floatfix]{revtex4-2}
\usepackage{graphicx}% Include figure files
\usepackage{dcolumn}% Align table columns on a decimal point
\usepackage{kotex}
\usepackage{bm}% bold math
\usepackage{braket} % Required for bra-ket notation
\usepackage{bbm}
\usepackage{amsmath}

\usepackage{algorithm}
\usepackage[noend]{algpseudocode}
\usepackage{subcaption}
\usepackage{wrapfig}
\usepackage{adjustbox}
\usepackage{relsize}
\usepackage{xcolor}
\usepackage{svg}
\usepackage{natbib}
\usepackage{multirow}
\usepackage{afterpage}
\usepackage{tikz} % required for circuit diagrams
\usetikzlibrary{quantikz} % required for circuit diagrams

\usepackage{hhline}
\usepackage{hyperref}

\definecolor{C1}{RGB}{52, 89, 149}
\definecolor{C2}{RGB}{251, 77, 61}
\definecolor{C3}{RGB}{3, 206, 164}
\definecolor{C4}{RGB}{202, 21, 81}
\usepackage{hyperref}
\hypersetup{colorlinks=true, linkcolor=C2, citecolor=C2, urlcolor=C1}

\usepackage[normalem]{ulem}
\usepackage{lipsum}

\newcommand{\beginsupplement}{
  \setcounter{table}{0}  
  \renewcommand{\thetable}{S\arabic{table}} 
  \setcounter{figure}{0} 
  \renewcommand{\thefigure}{S\arabic{figure}}
  \setcounter{section}{0}
  \setcounter{equation}{0}
  \renewcommand{\theequation}{S\arabic{equation}}
}

\makeatletter
\newcommand{\rightleftarrows}[2]{%
  \mathrel{\mathop{%
    \vcenter{\offinterlineskip\m@th
      \ialign{\hfil##\hfil\cr
        \hphantom{$\scriptstyle\mspace{8mu}{#1}\mspace{8mu}$}\cr
        \rightarrowfill\cr
        \vrule height0pt width 2em\cr
        \leftarrowfill\cr
        \hphantom{$\scriptstyle\mspace{8mu}{#2}\mspace{8mu}$}\cr
        \noalign{\kern-0.3ex}
      }%
    }%
  }\limits^{#1}_{#2}}%
}
\makeatother

\makeatletter
\newcommand{\rightarrows}[1]{%
  \mathrel{\mathop{%
    \vcenter{\offinterlineskip\m@th
      \ialign{\hfil##\hfil\cr
        \rightarrowfill\cr
        \hphantom{$\scriptstyle\mspace{8mu}{#1}\mspace{8mu}$}\cr
        \noalign{\kern-0.3ex}
      }%
    }%
  }\limits^{#1}}%
}
\makeatother

\newcommand{\Rom}[1]{\uppercase\expandafter{\romannumeral#1\relax}}

\begin{document}

\title{Magic State Injection on IBM Quantum Processors Above the Distillation Threshold}

% Authors & Institutions
\author{Younghun Kim}
\email{younghunk@student.unimelb.edu.au}
\affiliation{School of Physics, The University of Melbourne, Parkville, 3010, Victoria, Australia}
\affiliation{Data61, CSIRO, Clayton, 3168, Victoria, Australia}

\author{Martin Sevior}
\affiliation{School of Physics, The University of Melbourne, Parkville, 3010, Victoria, Australia}

\author{Muhammad Usman}
\email{muhammad.usman@unimelb.edu.au}
\affiliation{School of Physics, The University of Melbourne, Parkville, 3010, Victoria, Australia}
\affiliation{Data61, CSIRO, Clayton, 3168, Victoria, Australia}

\begin{abstract}
The surface code family is a promising approach to implementing fault-tolerant quantum computations. Universal fault-tolerance requires error-corrected non-Clifford operations, in addition to Clifford gates, and for the former, it is imperative to experimentally demonstrate additional resources known as magic states. Another challenge is to efficiently embed surface codes into quantum hardware with connectivity constraints. This work simultaneously addresses both challenges by employing a qubit-efficient rotated heavy-hexagonal surface code for IBM quantum processors (\texttt{ibm\_fez}) and implementing the magic state injection protocol. Our work reports error thresholds for both logical bit- and phase-flip errors, of $\approx0.37\%$ and $\approx0.31\%$, respectively, which are higher than the threshold values previously reported with traditional embedding. The post-selection-based preparation of logical magic states $|H_L\rangle$ and $|T_L\rangle$ achieve fidelities of $0.8806\pm0.0002$ and $0.8665\pm0.0003$, respectively, which are both above the magic state distillation threshold. Additionally, we report the minimum fidelity among injected arbitrary single logical qubit states as $0.8356\pm0.0003$. Our work demonstrates the potential for realising non-Clifford logical gates by producing high-fidelity logical magic states on IBM quantum devices.
\end{abstract}

\maketitle

\section{INTRODUCTION}
Quantum error correction is vital to achieving scalable and universal fault-tolerant quantum computation by suppressing inevitable errors in contemporary quantum computing \cite{shor_scheme_1995,calderbank_good_1996,steane_multiple_1996,terhal_quantum_2015}. Surface codes have emerged as one of the leading quantum error correction protocols, which promise to protect quantum information by encoding it across many entangled qubits \cite{gottesman_stabilizer_1997,bravyi_quantum_1998,dennis_topological_2002,kitaev_fault-tolerant_2003}. They stabilize qubits and focus errors into relatively easy-to-correct forms and require a straightforward coupling map \cite{fowler_towards_2012}. In recent years, surface code implementations, albeit on a small scale, have been demonstrated on a variety of quantum hardware platforms such as superconducting \cite{zhao_realization_2022,krinner_realizing_2022,google_quantum_ai_suppressing_2023,acharya_quantum_2024}, trapped ion systems \cite{berthusen_experiments_2024}, and neutral atom systems \cite{bluvstein_logical_2024}. However, research on surface code implementations on quantum processors remains at a preliminary stage and universal fault-tolerant quantum computing is still an open problem that requires demonstrations of a full surface code protected universal gate set including both Clifford and non-Clifford quantum operations \cite{reichardt_quantum_2005,litinski_game_2019,chamberland_universal_2022}.

Although the experimental relisation of Clifford gates with surface code formalism and its variants has been quite successful in the literature \cite{eastin_restrictions_2009,erhard_entangling_2021,ryan-anderson_implementing_2022,kim_transversal_2024,hetenyi_creating_2024,bluvstein_logical_2024,paetznick_demonstration_2024,ryan-anderson_high-fidelity_2024}, it is well known theoretically that the implementation of non-Clifford operations is a highly nontrivial task that requires employing a specific logical qubit state, known as a logical magic state \cite{bravyi_universal_2005,eastin_restrictions_2009}. Therefore, the experimental realization of the logical magic state is an important milestone towards achieving universal quantum computing \cite{egan_fault-tolerant_2021,postler_demonstration_2022,ye_logical_2023,gupta_encoding_2024}, which is a topic of ongoing research. Only two experimental studies \cite{ye_logical_2023, gupta_encoding_2024} have reported preparing magic states using surface codes. While Ref.~\cite{ye_logical_2023} has shown distance-3 implementations exceeding the distillation threshold value for square lattice code, the study in Ref. \cite{gupta_encoding_2024} was only distance-2 surface codes without scaling. In this work, we report the first distance-3 implementations of surface code-based magic state injection on an IBM quantum processor and demonstrate that any arbitrary single logical qubit state can be prepared, including logical magic states above the distillation threshold. Additionally, our work designs a rotated surface code embedding in the heavy-hexagonal architecture of IBM quantum processors. This is also the first such implementation on IBM devices and reduces the physical qubit requirements by approximately half compared to previous surface code embeddings \cite{mcewen_relaxing_2023,benito_comparative_2024}, thus significantly reducing the resource requirements for scalable surface code implementations.

\begin{figure*}[ht]
    \centering
    \includegraphics[width=0.8\textwidth]{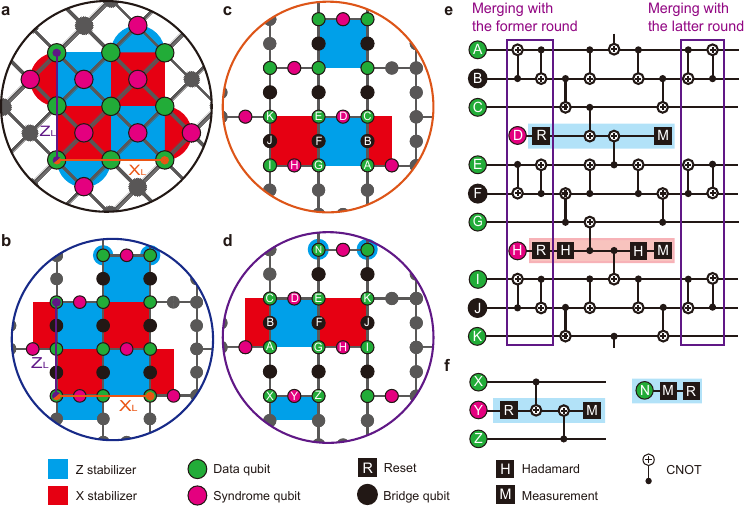}
    \caption{ \textbf{Embedding rotated surface code.} \textbf{a,b.} Qubit arrangements of the rotated surface code in the square and heavy-hexagon lattices. The logical Z and X operators are represented by the purple and orange solid lines, respectively.  \textbf{c,d.} The stabilizers of the code are divided into two sub-groups. \textbf{e.} A sub-round syndrome extraction circuit measures weight-four stabilizers including those at the side boundaries within the same sub-group. The first and last two layers of CNOT gates can be executed in parallel along with the former and latter sub-rounds. \textbf{f.} Subroutine circuits for measuring weights-two and -one Z stabilizers at top and bottom boundaries. }
    \label{fig1}
\end{figure*}

Our code exhibits a relatively high threshold compared to other lattice-compatible codes \cite{chamberland_topological_2020,kim_design_2023,mclauchlan_accommodating_2024,benito_comparative_2024}. We obtain $\approx0.37\%$ and $\approx0.31\%$ threshold values for both logical bit- and phase-flip errors, respectively. Additionally, we conduct magic state injection experiments using distance-3 codes on the \texttt{ibm\_fez} device. Our circuit can encode any arbitrary single-qubit state into the logical qubit state of the code. Among logically encoded states, the logical magic states relevant for the non-Clifford quantum operations namely $|H_L\rangle$ and $|T_L\rangle$ are prepared with fidelities of $0.8806\pm0.0002$ and $0.8665\pm0.0003$, respectively, both exceeding their distillation threshold values.

\section{Rotated Surface Code Embedding} \label{embeding_code}
The rotated surface code is an optimized variant of the conventional surface code that reduces resources while preserving code distance, by using a different orientation of qubits. Previous work has implemented surface codes in a heavy-hexagon lattice \cite{benito_comparative_2024}, but did not embed its rotated version. This work investigates the feasibility and efficiency of embedding the rotated surface code while respecting connectivity limitations. We estimate the threshold value for logical Pauli errors as a function of code distance and demonstrate encoding logical magic states in the rotated version of surface codes on one of IBM's devices.

In a square lattice, the rotated surface code displayed in Fig.\,\ref{fig1}a can concurrently stabilize data qubits in a subspace that yields +1 eigenvalues for its stabilizer group. The stabilizer group includes weight-four at the bulk and weight-two stabilizers at the boundaries, defined with commutable multi-qubit Pauli operators. However, the limited connectivity of the heavy-hexagon lattice makes embedding of rotated code a challenging task. This is because each qubit has at most three neighbors instead of four, which limits measuring weight-four stabilizers in the same manner as on the square lattice. The rotated surface code can be embedded in the heavy-hexagon lattice, as shown in Fig.\,\ref{fig1}b, by stabilizing data qubits through two consecutive sub-rounds. Each sub-round measures the corresponding subgroup stabilizers, as illustrated in Fig.\,\ref{fig3}c, and d. 

\begin{figure*}[t]
    \centering
    \includegraphics[width=0.95\textwidth]{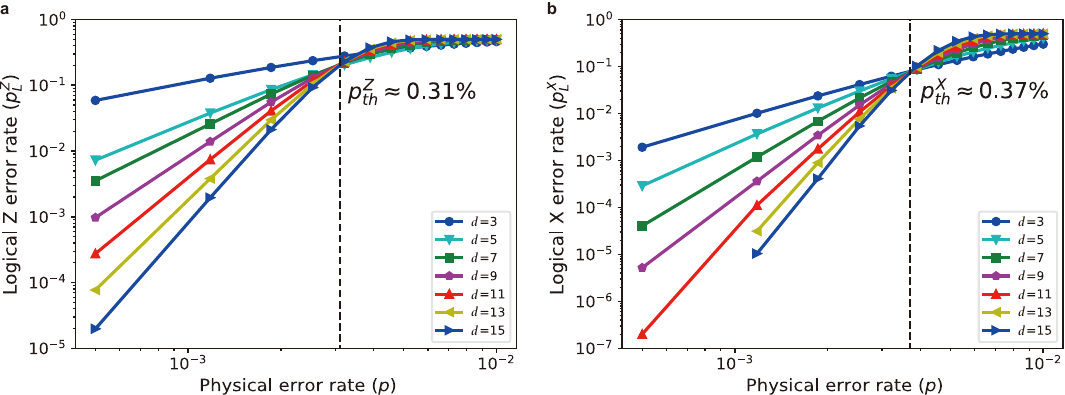}
    \caption{ \textbf{Logical error rates.} \textbf{a,b.} Logical Z and X error rates as functions of the code distance, denoted as $d$, and the physical error rate ($p$), showing error suppression via scaling code size. While the code distance ranges from 3 to 15, the physical error rate $p$ varies from $5\times10^{-4}$ to $10^{-2}$. The logical error rates are calculated using $5\times10^6$ samples. The threshold values are obtained as $p^Z_{\text{th}}\approx 0.31\%$ and $p^X_{\text{th}}\approx 0.37\%$ for logical Z and X errors, respectively. }
    \label{fig2}
\end{figure*}

\begin{subequations}
\begin{align}
\hat{Z}_A\hat{Z}_C\hat{Z}_E\hat{Z}_G &\rightleftarrows{\text{fold}}{\text{unfold}} \hat{Z}_C\hat{Z}_E \label{folding_a} \\ \hat{X}_E\hat{X}_G\hat{X}_I\hat{X}_K &\rightleftarrows{\text{fold}}{\text{unfold}} \hat{X}_G\hat{X}_I 
\label{folding_b}
\end{align}
\end{subequations}

The key idea behind measuring weight-four stabilizers, such as $\hat{Z}_A\hat{Z}_C\hat{Z}_E\hat{Z}_G$ and $\hat{X}_E\hat{X}_G\hat{X}_I\hat{X}_K$, while respecting the connectivity limitation is transforming them into weight-two stabilizers, in other words, folding stabilizers. The folded stabilizers are then measured using syndrome qubits to extract their eigenvalues. Following measurement, the original stabilizers are restored by reversing the series of quantum gates applied during the folding process, which is unfolding stabilizers. Fig.\,\ref{fig1}e shows a sub-round syndrome extraction circuit for the measurement of weight-four stabilizers following the process of (un)folding stabilizers as expressed in \eqref{folding_a} and \eqref{folding_b}. The circuit requires long-range interactions between data qubits, which can be done effectively by employing bridge qubits as intermediaries. To optimize the circuit, the first and last two layers of CNOT gates in the former and latter sub-rounds are applied simultaneously to minimize the circuit depth across sub-rounds.

While the stabilizers at the side boundaries can also be measured with those at the bulk, the stabilizers at the top or bottom boundaries need to be measured as weight-two and -one stabilizers as illustrated in Fig.\,\ref{fig1}f. For the Z stabilizers at the bottom boundary, the two-weight stabilizers can be directly measured using the syndrome qubits. For those at the top boundary, we update the stabilizer group by measuring weight-four or weight-one stabilizers, such as $\hat{Z}_N$, as sub-rounds executed alternatively. The measurement of the weight-one stabilizers collapses the corresponding data qubit states, allowing the remaining data qubits to span the code space where qubit information is protected from errors. Since the measurement of these stabilizers does not involve bridge qubits, they require relatively few time steps compared to those required for weight-four. However, we simultaneously apply measurements for stabilizers in the same sub-round regardless of their weight.

\begin{table}[h]
\centering
\caption{\textbf{Qubit requirements.} The required number of physical qubits for embedding rotated and unrotated surface codes on the heavy-hexagon lattice as a function of the code distance ($d$).}
\begin{tabular}{|| c | c | c ||}
\hline
            & Unrotated & Rotated\\
\hline
$\#$ of data qubits & $2d^2-1$   & $d^2+d-1$\\
\hline
$\#$ of qubits &  $5d^2 -2(d+1)$  & ${5/2}d^2  + 2d - {7/2}$\\
\hline
\end{tabular}
\label{qubit_num}
\end{table}

To evaluate the code, we computed logical error rates as a function of the physical error rates under the circuit-level uniform noise model (the details are in Methods). Fig.\,\ref{fig2}a and b show the logical error rates for the two states, either in the X ($|+_L\rangle$) basis or in Z ($|0_L\rangle$) basis, corresponding to the probability of logical Z and X errors, respectively. The logical error rates are calculated under various physical error rates ($p$) ranging from $5\times 10^{-4}$ to $10^{-2}$ and code distances ($d$) ranging from 3 to 15. Although the number of qubits required for the rotated code scales with the code distance as $O(d^2)$, the same as for its unrotated version, the former requires about half the number of qubits in the limit of large code distances. The number of qubits as a function of the code distance is listed in Table \ref{qubit_num}.

\begin{figure*}[t]
    \centering
    \includegraphics[width=\textwidth]{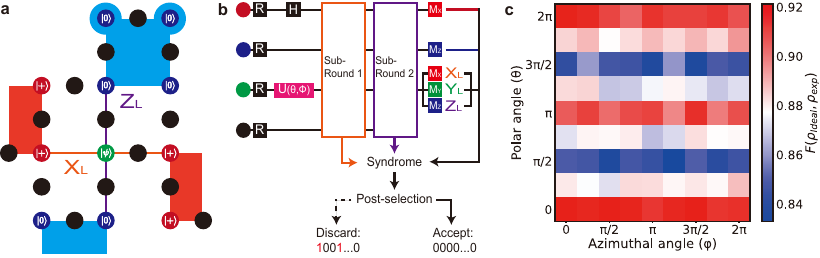}
    \caption{ \textbf{Magic state injection and implementation.} \textbf{a.} The initialization layout for magic state injection utilizing the embedded rotated surface code. Each data qubit is initialized with the ground state in either the Z basis ($\ket{0}$), represented as blue nodes, or the X basis ($\ket{+}$), represented as red nodes. The central data qubit, magic state, is prepared in the state $|\psi\rangle$ and depicted as a green node. \textbf{b.} The data qubits are initialized according to the layout, employing a single-qubit unitary operation (U) for the magic state. Following that, two consecutive sub-round syndrome extraction circuits are executed. The magic state is measured in the basis determined by the logical Pauli measurement, while the remaining data qubits are measured in the same basis as their initialization. The outcomes are used to detect errors and discard non-trivial syndromes through post-selection. \textbf{c.} The fidelities of raw logical magic states prepared using the circuit \textbf{b} are plotted on a plane with the polar ($\theta$) and azimuthal$ (\phi$) angles, ranging from 0 to $2\pi$. The values are estimated from the trivial syndromes obtained by sampling $2 \times 10^4$ times for logical measurements in each Pauli basis. }
    \label{fig3}
\end{figure*}

\begin{figure*}[t]
    \centering
    \includegraphics[width=0.6\textwidth]{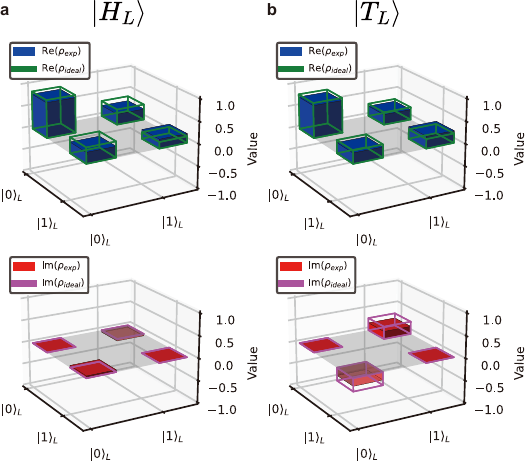}
    \caption{ \textbf{Density matrices of logical magic states.} \textbf{a,b.} The real and imaginary values of the density matrices for the ideal and experimental results are plotted for $|H_L\rangle$ and $|T_L\rangle$, with ideal values represented by lines and experiment values by bars. A grey plane indicates the region where the value is 0. }
    \label{fig4}
\end{figure*}

We obtain the threshold values for logical phase- and bit-flip errors as $\approx 0.31\%$ and $\approx 0.37\%$, respectively. Although its unrotated version has a threshold of $0.3\%$ \cite{benito_comparative_2024}, the highest among other lattice-compatible codes suggested so far, we find that the proposed code in this work achieves a slightly even higher threshold value for both error types. Notably, we observe a faster improvement in performance for bit-flip error corrections compared to phase-flip error corrections as the physical error rate decreases below the threshold, given the same code distance. Additionally, the threshold value for a logical Z error is smaller than that for a logical X error. This indicates an asymmetric feature, that has not been observed in the square lattice, in correcting the two types of errors under the unbiased noise model when rotating the embedded code in the heavy-hexagon structure. More details about the feature are discussed in Supplementary Section \ref{threshold_sec}.%\Rom{1}

\section{Magic State Injection Protocol} \label{magic_inject}

We next discuss a code-based magic state injection protocol. The protocol consists of four main steps: 1. Initialization, 2. Stabilizer measurement, 3. Logical Pauli measurement, and 4. Post-selection. 

\begin{enumerate}
  \item Initialization: Data qubits are prepared in their designated quantum states, as depicted in Fig.\,\ref{fig3}a. These states include the ground state of the Z basis ($|0\rangle$) or the X basis ($|+\rangle$). Additionally, the central qubit, the magic state, is prepared as $|\psi\rangle=\text{cos}(\theta/2)|0\rangle + e^{i\phi}\text{sin}(\theta/2)|1\rangle$, intended for injection at the logical qubit level, using a primitive single-qubit gate parameterized with $\theta$ and $\phi$. 
  
  \item Stabilizer measurement: Two sub-rounds of syndrome extraction circuits are performed and measure full stabilizers. 

  \item Logical Pauli measurement: As shown in Fig.\,\ref{fig3}b, the central data qubit is measured in the same Pauli basis as the interrogated logical Pauli measurement basis i.e. X, Y, or Z. In contrast, the remaining data qubits are measured in the same basis as initialized. 
  
  \item Post-selection: Based on the measured outcomes, we evaluated deterministic parity values described in \ref{item1} and \ref{item2}, to produce a syndrome. When there is an error, a non-trivial syndrome is created and therefore discarded, as shown in Fig.\,\ref{fig3}b. 
  
  \begin{enumerate}
      \item\label{item1} The measurement outcomes from the sub-round syndrome extraction circuits align with the initially conditioned stabilizers at the boundaries.
      \item\label{item2} The parity of the measurements for the data qubits associated with the boundary stabilizers matches the eigenvalue of the corresponding stabilizer.
  \end{enumerate}
\end{enumerate}

We decided to design the protocol using a distance-3 rotated surface code, which reduces the number of qubits by one-third compared to its unrotated version. The reduced qubit requirement minimizes errors and increases the success rate for the post-selection experiments. The data qubits are initialized to have eigenvalues of +1 at all boundary stabilizers. They are then stabilized through the sub-round syndrome extraction circuits. In the absence of noise, these processes map the data qubits into the codespace, forming a logical qubit. Depending on the state injected into the center data qubit using predefined parameters $\theta$ and $\phi$ in the initialization process, the state of the logical qubit is prepared as $|\psi_L\rangle=\text{cos}(\theta/2)|0_L\rangle + e^{i\phi}\text{sin}(\theta/2)|1_L\rangle$. When we measure the logical qubit on a logical Pauli basis, we estimate the measurement by the parity of the measured outcomes from the data qubits associated with the target basis, where $\hat{Y}_L=i\hat{X}_L\hat{Z}_L$ \cite{ye_logical_2023}. We can change the basis of this projective logical Pauli measurement by measuring the magic state on a target basis.

We conducted experiments using 25 physical qubits on \texttt{ibm\_fez} quantum device, one of IBM's quantum processors, accessed via the cloud. The experiments were conducted using the parameters $\theta$ and $\phi$ ranging from 0 to $2\pi$ in $\pi/4$ intervals. The details of the preparation of an arbitrary magic state ($|\psi\rangle$) through a primitive single-qubit gate are provided in Methods. Each logical Pauli measurement for the X, Y, and Z basis is repeated $2\times10^4$ times per state with samples taking approximately $6.8\mu$s. 

It is worth discussing the error rates of logic quantum gates, particularly two-qubit gates and hardware measurements. During our experiments, the average error rate from the two-qubit gates ($2.9\times 10^{-3}$) was below the threshold values for both types of code logical errors. In contrast, the same for the readout case ($1.6\times 10^{-2}$) was an order of magnitude higher compared to the threshold values. Further details of the error rates can be found in the Methods. Although this might make logical qubits more susceptible to measurement-induced logical errors than gate-induced ones, the post-selection in the experiments could effectively detect both measurement-induced and gate-induced errors during error detection, potentially improving the fidelity of logical magic states. As a result, the average acceptance rate for experiments for each logical Pauli measurement that met the post-selection criteria was $36.28\pm0.09\%$. In Methods, we show the acceptance rates of post-selection in eigenstates of logical Pauli operators.

\begin{equation}
F(\rho_{\text{ideal}},\rho_{\text{exp}}) = \left(Tr\left(\sqrt{\sqrt{\rho_{\text{exp}}}\rho_{\text{ideal}}\sqrt{\rho_{\text{exp}}}}\right)\right)^2
\label{fidelity}
\end{equation}

Based on the post-selected samples from the experiments on the IBM device, we calculate fidelities for logical qubit states. The fidelities are calculated between the ideal ($\rho_{\text{ideal}}$), and experimentally reconstructed density matrix ($\rho_{\text{exp}}$) as expressed in equation \eqref{fidelity}. While the ideal density matrix is obtained from the theoretical logical state of the qubit as $\rho_{\text{ideal}}=|\psi_L\rangle\langle\psi_L|$, the experimental density matrix is estimated based on the expectation values of the logical Pauli operators (see Methods). Furthermore, we compare theoretical and experimental logical Pauli expectation values in Supplementary Section \ref{expectation_sec}.%\Rom{2}

In Fig.\,\ref{fig3}c, we plot the fidelities for the experimentally encoded logical qubit states. When the polar angle ($\theta$) is an integer multiple of $\pi$ regardless of the azimuthal angle ($\phi$), logical qubits are prepared in the eigenstates of the logical Z operator where the values are prominently high. However, we observe a gradual decrease in the fidelities of qubit states, when the state is injected as the superposition of two computational Z basis states with different phases. Under the assumption that primitive gates in the hardware have no biased noise, we attribute the relative vulnerability of phase information to the inherent bias of the resilience of the code against bit-flip errors. This aligns with the asymmetric feature of the code analyzed theoretically. Furthermore, the minimum value has been found in the Y basis state with +1 eigenvalue from the logical Y operator as $0.8356\pm0.0003$, and the state is susceptible to both bit-flip and phase-flip errors. However, even with antisymmetric robustness against errors, we note that our protocol achieves an average fidelity of $0.882\pm0.006$. 

Finally, we test our model in the preparation of well-known logical magic states, $|H\rangle = \cos(\pi/8)|0\rangle + \sin(\pi/8)|1\rangle$ and $|T\rangle = \cos(\beta)|0\rangle + e^{i \pi/4}\sin(\beta)|1\rangle$, where $\cos(2\beta)=1/\sqrt{3}$. H- and T-type states can be used to realize phase-shift gates, which belong to non-Clifford gates \cite{bravyi_universal_2005}. The threshold fidelity values of $|H\rangle$, using a 7-to-1 distillation routine \cite{reichardt_quantum_2005}, and $|T\rangle$, using a 5-to-1 distillation routine \cite{bravyi_universal_2005}, are 0.854 and 0.827.

We conduct experiments to prepare the magic states and analyze their fidelities. The results are shown in Fig.\,\ref{fig4}a and b where we compare the ideal and experimental density matrices for the two logical magic states. The fidelity of the logical magic states, $|H_L\rangle$ and $|T_L\rangle$, are prepared with the fidelity $0.8806\pm0.0002$ and $0.8665\pm0.0003$, respectively, which are above the threshold for the distillation protocol. The uncertainty of fidelities is estimated using a bootstrapping technique (see Methods). 

\section{CONCLUSION AND DISCUSSION}
In this work, we prepare an arbitrary encoded logical single-qubit state using a rotated surface code on the \texttt{ibm\_fez} quantum processor. First, we demonstrate a rotated surface code, which requires around half the number of qubits as its unrotated version for large code distances in the heavy-hexagon structure of \texttt{ibm\_fez}. We compute the threshold values of the code and find an asymmetric feature in error correction. We have achieved high fidelity using the injection protocol on arbitrary single-qubit states for logical encoding. We also realize two logical magic states $|H_L\rangle$ and $|T_L\rangle$ type, which can be employed to implement non-Clifford gates for quantum error correction. The results show we exceed the threshold fidelity of magic state distillation on IBM hardware.

Several challenges remain to achieve the universality of the logical quantum gate set; however, our work marks an important step towards universal quantum computing by demonstrating the preparation of raw logical magic states on an IBM quantum device. In future work, it would be intriguing to conduct quantum memory experiments and demonstrate error suppression using our scheme, using code distances from 3 to 5, and even 7, by shaping the code as rectangular on 156 physical qubit devices. Furthermore, increasing the size of the logical magic states and implementing lattice surgery would be another promising avenue for future work \cite{li_magic_2015,horsman_surface_2012}. In summary, our protocols pave the way for promising near-term advancements in quantum error codes for quantum hardware with connectivity constraints such as the heavy-hexagon structure employed by the IBM quantum processors.

\section{METHODS}\label{method}

\textbf{Noise Model:} We evaluate threshold values of the rotated surface code embedded in the heavy-hexagonal structure under a circuit-level noise model. We adopt the noise model that decomposes error channels using Pauli operators. A depolarizing error channel is used, where the errors are not biased but have a uniform probabilistic distribution among the Pauli errors \cite{wang_threshold_2009}. When the error rate is $p$, the circuit-level noise model consists of the following noisy channels:

\begin{itemize}
  \item Single-Qubit Depolarizing Error Channel: A single qubit subjected to the error channel experiences Pauli errors ($\hat{X}$, $\hat{Y}$, and $\hat{Z}$) with equal probabilities. The error probabilities for $\hat{X}$, $\hat{Y}$, and $\hat{Z}$ are denoted by $p_X$, $p_Y$, and $p_Z$, respectively, satisfying the condition $p_X = p_Y = p_Z = p/3$. This occurs when a physical qubit is inactive and undergoing free evolution or when a single-qubit gate, such as $\hat{H}$, is applied.

  \item Initialization and Measurement Error Channel: A bit-flip error, with a probability of $p$, is applied before measurement ($M$) and after the reset gate ($R$) on the basis Z.

  \item Two-Qubit Depolarizing Error Channel: Two qubits are susceptible to Pauli errors when a two-qubit gate (CNOT) is applied. These Pauli errors of two qubits are represented by the set $\{\hat{X}, \hat{Y}, \hat{Z}, \hat{I}\}^{\otimes 2}/\{\hat{I} \otimes \hat{I}\}$. The probability of each error is uniform as $p/15$. 
\end{itemize}

\textbf{Calculating Threshold:} In this work, to compute a logical error rate, we turn measured results from syndrome extraction circuits into a syndrome to detect errors and use it to calculate a correction operator. Errors are detected through flipped measurement outcomes from the same syndrome qubits, which will produce ``1" bits in the syndrome, indicating the presence of errors between the circuit rounds for weight-four and -two stabilizers. However, results from weight-one stabilizers are directly used to detect errors. We use the open-source software tool Stim to generate syndrome samples using sub-round syndrome extraction circuits under circuit-level noise, decomposing noise as probabilistic Pauli gates \cite{gidney_stim_2021}. Furthermore, we use Pymatching to determine a correction operator as the most likely error based on the noise model \cite{higgott_pymatching_2022,higgott_sparse_2023}. A logical error rate is computed as the ratio of the average number of rounds to have a logical error for varying error rates and code distances. 

\textbf{IBM Hardware:} We conducted the experiments on 31st October 2024, utilizing 25 out of 156 physical qubits on \texttt{ibm\_fez} device. Measurement and two-qubit gate (CZ gate) error rates may vary per qubit within the device's configuration. The error rates for the chosen qubits in the experiments are the hardware's calibration data and are illustrated in Supplementary Fig.\,\ref{fig_app2}. The average error rates for readout and two-qubit gates were $1.6 \times 10^{-2}$ and $2.9 \times 10^{-3}$, respectively. %S2

\textbf{Implementation of the experiment:} We conducted optimized quantum circuits for the magic state injection protocol on the hardware. First, to optimize circuits, data qubits forming weight-one stabilizers are measured only once rather than twice, because there are redundant measurements of data qubits in the second sub-round syndrome extraction circuit and logical Pauli measurement. Second, we classically controlled the X gates based on the measurement results of the qubits to reset the physical qubits \cite{sundaresan_demonstrating_2023}. This approach reduced the cost of syndrome extraction circuits by minimizing the time to reset qubits. We applied dynamic decoupling to the physical qubits while inactive, to minimize unwanted perturbations during quantum operations, including measurements \cite{google_quantum_ai_exponential_2021}.

We used a U3 gate to rotate a single qubit to prepare the $|\psi\rangle$ state from the ground state ($|0\rangle$), where the gate takes three parameters $\theta$, $\phi$, and $\lambda$. The gate represents: 

\begin{equation}
U3(\theta,\phi,\lambda) = \begin{pmatrix} \text{cos}(\theta/2) & -e^{i\lambda}\text{sin}(\theta/2)\\ e^{i\phi}\text{sin}(\theta/2) & e^{i(\phi+\lambda)}\text{cos}(\theta/2)\end{pmatrix} \nonumber
\label{unitary}
\end{equation}

\noindent, where we have set $\lambda = 0$. We employed the Python library Qiskit to transpile circuits into the device's basis gates, enabling the execution of quantum circuits on the hardware \cite{qiskit}. 

\textbf{Qubit Tomography:} Any single-qubit density matrix ($\rho$) can be written in terms of its Pauli operators $\vec{\sigma}=(\hat{X}, \hat{Y}, \hat{Z})$ and the identity operator ($\mathbbm{1}$), such that:

\begin{equation}
\rho = {1\over{2}}(\mathbbm{1}+\vec{\sigma}\cdot\vec{r}) = \begin{pmatrix} 1-z & x-iy\\ x+iy & 1+z \end{pmatrix}
\end{equation}

\noindent, where $\vec{r} = (x, y, z)$ is a real vector representing the Bloch coordinates of $\rho$. The Bloch vector corresponds to the coefficients of each Pauli operator and can be used to reconstruct the density matrix. A single-qubit tomography is a process to estimate the Bloch vector based on the outcomes of non-commuting observables. The simplest method, direct inversion tomography, repeatedly measures a qubit state in the Pauli bases and obtains the expectation value for each basis, reconstructing the density matrix of the target single-qubit state \cite{schmied_quantum_2016}.

The numbers of repeated measurements along the logical Pauli X, Y, or Z basis can be denoted by $N_X$, $N_Y$, and $N_Z$, respectively. Each measurement yields one of two outcomes: ``up-state" with a +1 eigenvalue for the corresponding Pauli operator, or the ``down-state" corresponding to a $-1$ eigenvalue. The counts of the up- and down-states are represented as $N_{up}$ and $N_{down}$ and their sum is the total number of measurements for a particular Pauli basis. For example, $N_Z = N_{|0\rangle}+N_{|1\rangle}$, and similarly for the other Pauli bases, $N_X$ and $N_Y$. Based on these measured outcomes, the Bloch vector ($\vec{r}_{\text{exp}}$) can be estimated by the expectation value of each Pauli operator as follows \cite{schmied_quantum_2016}:

\begin{equation}
\vec{r}_{\text{exp}} = \left( {N_{|+\rangle}-N_{|-\rangle} \over {N_{X}}}, {N_{|+i\rangle}-N_{|-i\rangle} \over {N_{Y}}}, {N_{|0\rangle}-N_{|1\rangle} \over {N_{Z}}} \right)
\label{bloch}
\end{equation}

\begin{table}[h]
\centering
\caption{Acceptance rates of post-selection in different Pauli bases, where $N=2\times10^4$. $N_X$ is the number of samples passed the post-selection for measuring logical X measurement. Likewise for $N_Y$ and $N_Z$.}
\begin{tabular}{|| c | c | c | c ||}
\hline
\multirow{2}{*}{Logical state} & \multirow{2}{*}{$N_X/N$} & \multirow{2}{*}{$N_Y/N$} & \multirow{2}{*}{$N_Z/N$} \\
&&& \\
\hline
$|0\rangle_L$ & \multirow{2}{*}{0.3643}   & \multirow{2}{*}{0.3704}   & \multirow{2}{*}{0.3619}  \\
($\theta=0,\phi=0$)&&& \\
\hline
$|1\rangle_L$ & \multirow{2}{*}{0.3643}   & \multirow{2}{*}{0.3685}   & \multirow{2}{*}{0.3548}  \\
($\theta=\pi,\phi=0$)&&& \\
\hline
$|+\rangle_L$ & \multirow{2}{*}{0.3581}   & \multirow{2}{*}{0.3615}   & \multirow{2}{*}{0.3511} \\
($\theta=\pi/2,\phi=0$)&&& \\
\hline
$|-\rangle_L$ & \multirow{2}{*}{0.3713}   & \multirow{2}{*}{0.3674}   & \multirow{2}{*}{0.3677} \\
($\theta=\pi/2,\phi=\pi$)&&& \\
\hline
$|+i\rangle_L$ & \multirow{2}{*}{0.3616}   & \multirow{2}{*}{0.3715}  & \multirow{2}{*}{0.3633} \\
($\theta=\pi/2,\phi=\pi/2$)&&& \\
\hline
$|-i\rangle_L$ & \multirow{2}{*}{0.3669}   & \multirow{2}{*}{0.3734}  & \multirow{2}{*}{0.3613}  \\
($\theta=\pi/2,\phi=3\pi/2$)&&& \\
\hline
\end{tabular}
\label{acceptance}
\end{table}

As mentioned in the main paper, logical Pauli measurements are repeated $2\times10^4$ per injected magic state. We discard any sample which has a non-trivial syndrome, it may vary the number of samples for $N_{up}$ and $N_{down}$ states. The acceptance rates of experimental results for the eigenstates of Pauli operators are listed in Table \ref{acceptance}. The samples that passed the post-selection are then used to calculate expectation values of logical Pauli operators corresponding to logical magic states.

\textbf{Bootstrapping:} The confidence intervals in experimental data are estimated using a bootstrapping method \cite{bootstrap}. We classically resampled using the probability distribution obtained from the experiments.

\section*{ACKNOWLEDGMENT}
YK acknowledges the support of the CSIRO Research Training Program Scholarship and the University of Melbourne Research Training Scholarship. The University of Melbourne supported the research through the establishment of the IBM Quantum Network Hub.

\section*{Data Availability}
All datasets are available in the manuscript figures. Further data and source code can be made available upon reasonable request to the corresponding authors. 

\section*{Author Contributions}
YK developed and implemented the rotated surface code and magic state injection protocols under the supervision of MU and MS. YK carried out all experiments and plotted figures with input from MU. All authors discussed and analyzed the data. YK wrote the manuscript with input from MU and MS.

% \bibliographystyle{naturemag}
% \bibliography{References.bib}

\begin{thebibliography}{0}%
\makeatletter
\providecommand \@ifxundefined [1]{%
 \@ifx{#1\undefined}
}%
\providecommand \@ifnum [1]{%
 \ifnum #1\expandafter \@firstoftwo
 \else \expandafter \@secondoftwo
 \fi
}%
\providecommand \@ifx [1]{%
 \ifx #1\expandafter \@firstoftwo
 \else \expandafter \@secondoftwo
 \fi
}%
\providecommand \natexlab [1]{#1}%
\providecommand \enquote  [1]{``#1''}%
\providecommand \bibnamefont  [1]{#1}%
\providecommand \bibfnamefont [1]{#1}%
\providecommand \citenamefont [1]{#1}%
\providecommand \href@noop [0]{\@secondoftwo}%
\providecommand \href [0]{\begingroup \@sanitize@url \@href}%
\providecommand \@href[1]{\@@startlink{#1}\@@href}%
\providecommand \@@href[1]{\endgroup#1\@@endlink}%
\providecommand \@sanitize@url [0]{\catcode `\\12\catcode `\$12\catcode `\&12\catcode `\#12\catcode `\^12\catcode `\_12\catcode `\%12\relax}%
\providecommand \@@startlink[1]{}%
\providecommand \@@endlink[0]{}%
\providecommand \url  [0]{\begingroup\@sanitize@url \@url }%
\providecommand \@url [1]{\endgroup\@href {#1}{\urlprefix }}%
\providecommand \urlprefix  [0]{URL }%
\providecommand \Eprint [0]{\href }%
\providecommand \doibase [0]{https://doi.org/}%
\providecommand \selectlanguage [0]{\@gobble}%
\providecommand \bibinfo  [0]{\@secondoftwo}%
\providecommand \bibfield  [0]{\@secondoftwo}%
\providecommand \translation [1]{[#1]}%
\providecommand \BibitemOpen [0]{}%
\providecommand \bibitemStop [0]{}%
\providecommand \bibitemNoStop [0]{.\EOS\space}%
\providecommand \EOS [0]{\spacefactor3000\relax}%
\providecommand \BibitemShut  [1]{\csname bibitem#1\endcsname}%
\let\auto@bib@innerbib\@empty
%</preamble>
\end{thebibliography}%


\begin{thebibliography}{99}
\expandafter\ifx\csname url\endcsname\relax
  \def\url#1{\texttt{#1}}\fi
\expandafter\ifx\csname urlprefix\endcsname\relax\def\urlprefix{URL }\fi
\providecommand{\bibinfo}[2]{#2}
\providecommand{\eprint}[2][]{\url{#2}}

\bibitem{shor_scheme_1995}
\bibinfo{author}{Shor, P.~W.}
\newblock \bibinfo{title}{Scheme for reducing decoherence in quantum computer memory}.
\newblock \emph{\bibinfo{journal}{Physical Review A}} \textbf{\bibinfo{volume}{52}}, \bibinfo{pages}{R2493--R2496} (\bibinfo{year}{1995}).
\newblock \urlprefix\url{https://link.aps.org/doi/10.1103/PhysRevA.52.R2493}.

\bibitem{calderbank_good_1996}
\bibinfo{author}{Calderbank, A.~R.} \& \bibinfo{author}{Shor, P.~W.}
\newblock \bibinfo{title}{Good quantum error-correcting codes exist}.
\newblock \emph{\bibinfo{journal}{Physical Review A}} \textbf{\bibinfo{volume}{54}}, \bibinfo{pages}{1098--1105} (\bibinfo{year}{1996}).
\newblock \urlprefix\url{https://link.aps.org/doi/10.1103/PhysRevA.54.1098}.

\bibitem{steane_multiple_1996}
\bibinfo{author}{Steane, A.}
\newblock \bibinfo{title}{Multiple {Particle} {Interference} and {Quantum} {Error} {Correction}}.
\newblock \emph{\bibinfo{journal}{Proceedings of the Royal Society of London. Series A: Mathematical, Physical and Engineering Sciences}} \textbf{\bibinfo{volume}{452}}, \bibinfo{pages}{2551--2577} (\bibinfo{year}{1996}).
\newblock \urlprefix\url{http://arxiv.org/abs/quant-ph/9601029}.
\newblock \bibinfo{note}{ArXiv:quant-ph/9601029}.

\bibitem{terhal_quantum_2015}
\bibinfo{author}{Terhal, B.~M.}
\newblock \bibinfo{title}{Quantum error correction for quantum memories}.
\newblock \emph{\bibinfo{journal}{Reviews of Modern Physics}} \textbf{\bibinfo{volume}{87}}, \bibinfo{pages}{307--346} (\bibinfo{year}{2015}).
\newblock \urlprefix\url{https://link.aps.org/doi/10.1103/RevModPhys.87.307}.

\bibitem{gottesman_stabilizer_1997}
\bibinfo{author}{Gottesman, D.}
\newblock \bibinfo{title}{Stabilizer {Codes} and {Quantum} {Error} {Correction}} (\bibinfo{year}{1997}).
\newblock \urlprefix\url{http://arxiv.org/abs/quant-ph/9705052}.
\newblock \bibinfo{note}{ArXiv:quant-ph/9705052}.

\bibitem{bravyi_quantum_1998}
\bibinfo{author}{Bravyi, S.~B.} \& \bibinfo{author}{Kitaev, A.~Y.}
\newblock \bibinfo{title}{Quantum codes on a lattice with boundary} (\bibinfo{year}{1998}).
\newblock \urlprefix\url{http://arxiv.org/abs/quant-ph/9811052}.
\newblock \bibinfo{note}{ArXiv:quant-ph/9811052}.

\bibitem{dennis_topological_2002}
\bibinfo{author}{Dennis, E.}, \bibinfo{author}{Kitaev, A.}, \bibinfo{author}{Landahl, A.} \& \bibinfo{author}{Preskill, J.}
\newblock \bibinfo{title}{Topological quantum memory}.
\newblock \emph{\bibinfo{journal}{Journal of Mathematical Physics}} \textbf{\bibinfo{volume}{43}}, \bibinfo{pages}{4452--4505} (\bibinfo{year}{2002}).
\newblock \urlprefix\url{http://arxiv.org/abs/quant-ph/0110143}.
\newblock \bibinfo{note}{ArXiv:quant-ph/0110143}.

\bibitem{kitaev_fault-tolerant_2003}
\bibinfo{author}{Kitaev, A.}
\newblock \bibinfo{title}{Fault-tolerant quantum computation by anyons}.
\newblock \emph{\bibinfo{journal}{Annals of Physics}} \textbf{\bibinfo{volume}{303}}, \bibinfo{pages}{2--30} (\bibinfo{year}{2003}).
\newblock \urlprefix\url{https://linkinghub.elsevier.com/retrieve/pii/S0003491602000180}.

\bibitem{fowler_towards_2012}
\bibinfo{author}{Fowler, A.~G.}, \bibinfo{author}{Whiteside, A.~C.} \& \bibinfo{author}{Hollenberg, L. C.~L.}
\newblock \bibinfo{title}{Towards {Practical} {Classical} {Processing} for the {Surface} {Code}}.
\newblock \emph{\bibinfo{journal}{Physical Review Letters}} \textbf{\bibinfo{volume}{108}}, \bibinfo{pages}{180501} (\bibinfo{year}{2012}).
\newblock \urlprefix\url{https://link.aps.org/doi/10.1103/PhysRevLett.108.180501}.

\bibitem{zhao_realization_2022}
\bibinfo{author}{Zhao, Y.} \emph{et~al.}
\newblock \bibinfo{title}{Realization of an {Error}-{Correcting} {Surface} {Code} with {Superconducting} {Qubits}}.
\newblock \emph{\bibinfo{journal}{Physical Review Letters}} \textbf{\bibinfo{volume}{129}}, \bibinfo{pages}{030501} (\bibinfo{year}{2022}).
\newblock \urlprefix\url{https://link.aps.org/doi/10.1103/PhysRevLett.129.030501}.

\bibitem{krinner_realizing_2022}
\bibinfo{author}{Krinner, S.} \emph{et~al.}
\newblock \bibinfo{title}{Realizing repeated quantum error correction in a distance-three surface code}.
\newblock \emph{\bibinfo{journal}{Nature}} \textbf{\bibinfo{volume}{605}}, \bibinfo{pages}{669--674} (\bibinfo{year}{2022}).
\newblock \urlprefix\url{https://www.nature.com/articles/s41586-022-04566-8}.

\bibitem{google_quantum_ai_suppressing_2023}
\bibinfo{author}{{Google Quantum AI}} \emph{et~al.}
\newblock \bibinfo{title}{Suppressing quantum errors by scaling a surface code logical qubit}.
\newblock \emph{\bibinfo{journal}{Nature}} \textbf{\bibinfo{volume}{614}}, \bibinfo{pages}{676--681} (\bibinfo{year}{2023}).
\newblock \urlprefix\url{https://www.nature.com/articles/s41586-022-05434-1}.

\bibitem{acharya_quantum_2024}
\bibinfo{author}{Acharya, R.} \emph{et~al.}
\newblock \bibinfo{title}{Quantum error correction below the surface code threshold} (\bibinfo{year}{2024}).
\newblock \urlprefix\url{http://arxiv.org/abs/2408.13687}.
\newblock \bibinfo{note}{ArXiv:2408.13687 [quant-ph]}.

\bibitem{berthusen_experiments_2024}
\bibinfo{author}{Berthusen, N.} \emph{et~al.}
\newblock \bibinfo{title}{Experiments with the {4D} {Surface} {Code} on a {QCCD} {Quantum} {Computer}} (\bibinfo{year}{2024}).
\newblock \urlprefix\url{http://arxiv.org/abs/2408.08865}.
\newblock \bibinfo{note}{ArXiv:2408.08865 [quant-ph]}.

\bibitem{bluvstein_logical_2024}
\bibinfo{author}{Bluvstein, D.} \emph{et~al.}
\newblock \bibinfo{title}{Logical quantum processor based on reconfigurable atom arrays}.
\newblock \emph{\bibinfo{journal}{Nature}} \textbf{\bibinfo{volume}{626}}, \bibinfo{pages}{58--65} (\bibinfo{year}{2024}).
\newblock \urlprefix\url{https://www.nature.com/articles/s41586-023-06927-3}.

\bibitem{reichardt_quantum_2005}
\bibinfo{author}{Reichardt, B.~W.}
\newblock \bibinfo{title}{Quantum {Universality} from {Magic} {States} {Distillation} {Applied} to {CSS} {Codes}}.
\newblock \emph{\bibinfo{journal}{Quantum Information Processing}} \textbf{\bibinfo{volume}{4}}, \bibinfo{pages}{251--264} (\bibinfo{year}{2005}).
\newblock \urlprefix\url{http://link.springer.com/10.1007/s11128-005-7654-8}.

\bibitem{litinski_game_2019}
\bibinfo{author}{Litinski, D.}
\newblock \bibinfo{title}{A {Game} of {Surface} {Codes}: {Large}-{Scale} {Quantum} {Computing} with {Lattice} {Surgery}}.
\newblock \emph{\bibinfo{journal}{Quantum}} \textbf{\bibinfo{volume}{3}}, \bibinfo{pages}{128} (\bibinfo{year}{2019}).
\newblock \urlprefix\url{http://arxiv.org/abs/1808.02892}.
\newblock \bibinfo{note}{ArXiv:1808.02892 [quant-ph]}.

\bibitem{chamberland_universal_2022}
\bibinfo{author}{Chamberland, C.} \& \bibinfo{author}{Campbell, E.~T.}
\newblock \bibinfo{title}{Universal {Quantum} {Computing} with {Twist}-{Free} and {Temporally} {Encoded} {Lattice} {Surgery}}.
\newblock \emph{\bibinfo{journal}{PRX Quantum}} \textbf{\bibinfo{volume}{3}}, \bibinfo{pages}{010331} (\bibinfo{year}{2022}).
\newblock \urlprefix\url{https://link.aps.org/doi/10.1103/PRXQuantum.3.010331}.

\bibitem{eastin_restrictions_2009}
\bibinfo{author}{Eastin, B.} \& \bibinfo{author}{Knill, E.}
\newblock \bibinfo{title}{Restrictions on {Transversal} {Encoded} {Quantum} {Gate} {Sets}}.
\newblock \emph{\bibinfo{journal}{Physical Review Letters}} \textbf{\bibinfo{volume}{102}}, \bibinfo{pages}{110502} (\bibinfo{year}{2009}).
\newblock \urlprefix\url{https://link.aps.org/doi/10.1103/PhysRevLett.102.110502}.

\bibitem{erhard_entangling_2021}
\bibinfo{author}{Erhard, A.} \emph{et~al.}
\newblock \bibinfo{title}{Entangling logical qubits with lattice surgery}.
\newblock \emph{\bibinfo{journal}{Nature}} \textbf{\bibinfo{volume}{589}}, \bibinfo{pages}{220--224} (\bibinfo{year}{2021}).
\newblock \urlprefix\url{http://arxiv.org/abs/2006.03071}.
\newblock \bibinfo{note}{ArXiv:2006.03071 [quant-ph]}.

\bibitem{ryan-anderson_implementing_2022}
\bibinfo{author}{Ryan-Anderson, C.} \emph{et~al.}
\newblock \bibinfo{title}{Implementing {Fault}-tolerant {Entangling} {Gates} on the {Five}-qubit {Code} and the {Color} {Code}} (\bibinfo{year}{2022}).
\newblock \urlprefix\url{http://arxiv.org/abs/2208.01863}.
\newblock \bibinfo{note}{ArXiv:2208.01863 [quant-ph]}.

\bibitem{kim_transversal_2024}
\bibinfo{author}{Kim, Y.}, \bibinfo{author}{Sevior, M.} \& \bibinfo{author}{Usman, M.}
\newblock \bibinfo{title}{Transversal {CNOT} gate with multi-cycle error correction} (\bibinfo{year}{2024}).
\newblock \urlprefix\url{http://arxiv.org/abs/2406.12267}.
\newblock \bibinfo{note}{ArXiv:2406.12267 [quant-ph]}.

\bibitem{hetenyi_creating_2024}
\bibinfo{author}{Hetényi, B.} \& \bibinfo{author}{Wootton, J.~R.}
\newblock \bibinfo{title}{Creating entangled logical qubits in the heavy-hex lattice with topological codes} (\bibinfo{year}{2024}).
\newblock \urlprefix\url{http://arxiv.org/abs/2404.15989}.
\newblock \bibinfo{note}{ArXiv:2404.15989 [quant-ph]}.

\bibitem{paetznick_demonstration_2024}
\bibinfo{author}{Paetznick, A.} \emph{et~al.}
\newblock \bibinfo{title}{Demonstration of logical qubits and repeated error correction with better-than-physical error rates} (\bibinfo{year}{2024}).
\newblock \urlprefix\url{http://arxiv.org/abs/2404.02280}.
\newblock \bibinfo{note}{ArXiv:2404.02280 [quant-ph]}.

\bibitem{ryan-anderson_high-fidelity_2024}
\bibinfo{author}{Ryan-Anderson, C.} \emph{et~al.}
\newblock \bibinfo{title}{High-fidelity and {Fault}-tolerant {Teleportation} of a {Logical} {Qubit} using {Transversal} {Gates} and {Lattice} {Surgery} on a {Trapped}-ion {Quantum} {Computer}} (\bibinfo{year}{2024}).
\newblock \urlprefix\url{http://arxiv.org/abs/2404.16728}.
\newblock \bibinfo{note}{ArXiv:2404.16728 [quant-ph]}.

\bibitem{bravyi_universal_2005}
\bibinfo{author}{Bravyi, S.} \& \bibinfo{author}{Kitaev, A.}
\newblock \bibinfo{title}{Universal quantum computation with ideal {Clifford} gates and noisy ancillas}.
\newblock \emph{\bibinfo{journal}{Physical Review A}} \textbf{\bibinfo{volume}{71}}, \bibinfo{pages}{022316} (\bibinfo{year}{2005}).
\newblock \urlprefix\url{https://link.aps.org/doi/10.1103/PhysRevA.71.022316}.

\bibitem{egan_fault-tolerant_2021}
\bibinfo{author}{Egan, L.} \emph{et~al.}
\newblock \bibinfo{title}{Fault-tolerant control of an error-corrected qubit}.
\newblock \emph{\bibinfo{journal}{Nature}} \textbf{\bibinfo{volume}{598}}, \bibinfo{pages}{281--286} (\bibinfo{year}{2021}).
\newblock \urlprefix\url{https://www.nature.com/articles/s41586-021-03928-y}.

\bibitem{postler_demonstration_2022}
\bibinfo{author}{Postler, L.} \emph{et~al.}
\newblock \bibinfo{title}{Demonstration of fault-tolerant universal quantum gate operations}.
\newblock \emph{\bibinfo{journal}{Nature}} \textbf{\bibinfo{volume}{605}}, \bibinfo{pages}{675--680} (\bibinfo{year}{2022}).
\newblock \urlprefix\url{https://www.nature.com/articles/s41586-022-04721-1}.

\bibitem{ye_logical_2023}
\bibinfo{author}{Ye, Y.} \emph{et~al.}
\newblock \bibinfo{title}{Logical {Magic} {State} {Preparation} with {Fidelity} beyond the {Distillation} {Threshold} on a {Superconducting} {Quantum} {Processor}}.
\newblock \emph{\bibinfo{journal}{Physical Review Letters}} \textbf{\bibinfo{volume}{131}}, \bibinfo{pages}{210603} (\bibinfo{year}{2023}).
\newblock \urlprefix\url{https://link.aps.org/doi/10.1103/PhysRevLett.131.210603}.

\bibitem{gupta_encoding_2024}
\bibinfo{author}{Gupta, R.~S.} \emph{et~al.}
\newblock \bibinfo{title}{Encoding a magic state with beyond break-even fidelity}.
\newblock \emph{\bibinfo{journal}{Nature}} \textbf{\bibinfo{volume}{625}}, \bibinfo{pages}{259--263} (\bibinfo{year}{2024}).
\newblock \urlprefix\url{https://www.nature.com/articles/s41586-023-06846-3}.

\bibitem{mcewen_relaxing_2023}
\bibinfo{author}{McEwen, M.}, \bibinfo{author}{Bacon, D.} \& \bibinfo{author}{Gidney, C.}
\newblock \bibinfo{title}{Relaxing {Hardware} {Requirements} for {Surface} {Code} {Circuits} using {Time}-dynamics}.
\newblock \emph{\bibinfo{journal}{Quantum}} \textbf{\bibinfo{volume}{7}}, \bibinfo{pages}{1172} (\bibinfo{year}{2023}).
\newblock \urlprefix\url{http://arxiv.org/abs/2302.02192}.
\newblock \bibinfo{note}{ArXiv:2302.02192 [quant-ph]}.

\bibitem{benito_comparative_2024}
\bibinfo{author}{Benito, C.}, \bibinfo{author}{López, E.}, \bibinfo{author}{Peropadre, B.} \& \bibinfo{author}{Bermudez, A.}
\newblock \bibinfo{title}{Comparative study of quantum error correction strategies for the heavy-hexagonal lattice} (\bibinfo{year}{2024}).
\newblock \urlprefix\url{http://arxiv.org/abs/2402.02185}.
\newblock \bibinfo{note}{ArXiv:2402.02185 [quant-ph]}.

\bibitem{chamberland_topological_2020}
\bibinfo{author}{Chamberland, C.}, \bibinfo{author}{Zhu, G.}, \bibinfo{author}{Yoder, T.~J.}, \bibinfo{author}{Hertzberg, J.~B.} \& \bibinfo{author}{Cross, A.~W.}
\newblock \bibinfo{title}{Topological and {Subsystem} {Codes} on {Low}-{Degree} {Graphs} with {Flag} {Qubits}}.
\newblock \emph{\bibinfo{journal}{Physical Review X}} \textbf{\bibinfo{volume}{10}}, \bibinfo{pages}{011022} (\bibinfo{year}{2020}).
\newblock \urlprefix\url{https://link.aps.org/doi/10.1103/PhysRevX.10.011022}.

\bibitem{kim_design_2023}
\bibinfo{author}{Kim, Y.}, \bibinfo{author}{Kang, J.} \& \bibinfo{author}{Kwon, Y.}
\newblock \bibinfo{title}{Design of quantum error correcting code for biased error on heavy-hexagon structure}.
\newblock \emph{\bibinfo{journal}{Quantum Information Processing}} \textbf{\bibinfo{volume}{22}}, \bibinfo{pages}{230} (\bibinfo{year}{2023}).
\newblock \urlprefix\url{https://link.springer.com/10.1007/s11128-023-03979-2}.

\bibitem{mclauchlan_accommodating_2024}
\bibinfo{author}{McLauchlan, C.}, \bibinfo{author}{Gehér, G.~P.} \& \bibinfo{author}{Moylett, A.~E.}
\newblock \bibinfo{title}{Accommodating {Fabrication} {Defects} on {Floquet} {Codes} with {Minimal} {Hardware} {Requirements}} (\bibinfo{year}{2024}).
\newblock \urlprefix\url{http://arxiv.org/abs/2405.15854}.
\newblock \bibinfo{note}{ArXiv:2405.15854 [quant-ph]}.

\bibitem{li_magic_2015}
\bibinfo{author}{Li, Y.}
\newblock \bibinfo{title}{A magic state’s fidelity can be superior to the operations that created it}.
\newblock \emph{\bibinfo{journal}{New Journal of Physics}} \textbf{\bibinfo{volume}{17}}, \bibinfo{pages}{023037} (\bibinfo{year}{2015}).
\newblock \urlprefix\url{https://iopscience.iop.org/article/10.1088/1367-2630/17/2/023037}.

\bibitem{horsman_surface_2012}
\bibinfo{author}{Horsman, D.}, \bibinfo{author}{Fowler, A.~G.}, \bibinfo{author}{Devitt, S.} \& \bibinfo{author}{Meter, R.~V.}
\newblock \bibinfo{title}{Surface code quantum computing by lattice surgery}.
\newblock \emph{\bibinfo{journal}{New Journal of Physics}} \textbf{\bibinfo{volume}{14}}, \bibinfo{pages}{123011} (\bibinfo{year}{2012}).
\newblock \urlprefix\url{https://iopscience.iop.org/article/10.1088/1367-2630/14/12/123011}.

\bibitem{wang_threshold_2009}
\bibinfo{author}{Wang, D.~S.}, \bibinfo{author}{Fowler, A.~G.}, \bibinfo{author}{Stephens, A.~M.} \& \bibinfo{author}{Hollenberg, L. C.~L.}
\newblock \bibinfo{title}{Threshold error rates for the toric and surface codes} (\bibinfo{year}{2009}).
\newblock \urlprefix\url{http://arxiv.org/abs/0905.0531}.
\newblock \bibinfo{note}{ArXiv:0905.0531 [quant-ph]}.

\bibitem{gidney_stim_2021}
\bibinfo{author}{Gidney, C.}
\newblock \bibinfo{title}{Stim: a fast stabilizer circuit simulator}.
\newblock \emph{\bibinfo{journal}{Quantum}} \textbf{\bibinfo{volume}{5}}, \bibinfo{pages}{497} (\bibinfo{year}{2021}).
\newblock \urlprefix\url{http://arxiv.org/abs/2103.02202}.
\newblock \bibinfo{note}{ArXiv:2103.02202 [quant-ph]}.

\bibitem{higgott_pymatching_2022}
\bibinfo{author}{Higgott, O.}
\newblock \bibinfo{title}{{PyMatching}: {A} {Python} {Package} for {Decoding} {Quantum} {Codes} with {Minimum}-{Weight} {Perfect} {Matching}}.
\newblock \emph{\bibinfo{journal}{ACM Transactions on Quantum Computing}} \textbf{\bibinfo{volume}{3}}, \bibinfo{pages}{1--16} (\bibinfo{year}{2022}).
\newblock \urlprefix\url{https://dl.acm.org/doi/10.1145/3505637}.

\bibitem{higgott_sparse_2023}
\bibinfo{author}{Higgott, O.} \& \bibinfo{author}{Gidney, C.}
\newblock \bibinfo{title}{Sparse {Blossom}: correcting a million errors per core second with minimum-weight matching} (\bibinfo{year}{2023}).
\newblock \urlprefix\url{http://arxiv.org/abs/2303.15933}.
\newblock \bibinfo{note}{ArXiv:2303.15933 [quant-ph]}.

\bibitem{sundaresan_demonstrating_2023}
\bibinfo{author}{Sundaresan, N.} \emph{et~al.}
\newblock \bibinfo{title}{Demonstrating multi-round subsystem quantum error correction using matching and maximum likelihood decoders}.
\newblock \emph{\bibinfo{journal}{Nature Communications}} \textbf{\bibinfo{volume}{14}}, \bibinfo{pages}{2852} (\bibinfo{year}{2023}).
\newblock \urlprefix\url{https://www.nature.com/articles/s41467-023-38247-5}.

\bibitem{google_quantum_ai_exponential_2021}
\bibinfo{author}{{Google Quantum AI}} \emph{et~al.}
\newblock \bibinfo{title}{Exponential suppression of bit or phase errors with cyclic error correction}.
\newblock \emph{\bibinfo{journal}{Nature}} \textbf{\bibinfo{volume}{595}}, \bibinfo{pages}{383--387} (\bibinfo{year}{2021}).
\newblock \urlprefix\url{https://www.nature.com/articles/s41586-021-03588-y}.

\bibitem{qiskit}
\bibinfo{author}{{IBM Quantum, and Community}}.
\newblock \emph{\bibinfo{title}{Qiskit: An open-source framework for quantum computing}} (\bibinfo{year}{2021}).
\newblock \urlprefix\url{https://doi.org/10.5281/zenodo.2573505}.

\bibitem{schmied_quantum_2016}
\bibinfo{author}{Schmied, R.}
\newblock \bibinfo{title}{Quantum state tomography of a single qubit: comparison of methods}.
\newblock \emph{\bibinfo{journal}{Journal of Modern Optics}} \textbf{\bibinfo{volume}{63}}, \bibinfo{pages}{1744--1758} (\bibinfo{year}{2016}).
\newblock \urlprefix\url{https://www.tandfonline.com/doi/full/10.1080/09500340.2016.1142018}.

\bibitem{bootstrap}
\bibinfo{author}{B., E.} \& \bibinfo{author}{R.J, T.}
\newblock \emph{\bibinfo{title}{An Introduction to the Bootstrap}} (\bibinfo{publisher}{Chapman and Hall/CRC}, \bibinfo{year}{1994}).
\newblock \urlprefix\url{https://doi.org/10.1201/9780429246593}.

\expandafter\ifx\csname url\endcsname\relax
  \def\url#1{\texttt{#1}}\fi
\expandafter\ifx\csname urlprefix\endcsname\relax\def\urlprefix{URL }\fi
\providecommand{\bibinfo}[2]{#2}
\providecommand{\eprint}[2][]{\url{#2}}

\bibitem{google_quantum_ai_exponential_2021}
\bibinfo{author}{{Google Quantum AI}} \emph{et~al.}
\newblock \bibinfo{title}{Exponential suppression of bit or phase errors with cyclic error correction}.
\newblock \emph{\bibinfo{journal}{Nature}} \textbf{\bibinfo{volume}{595}}, \bibinfo{pages}{383--387} (\bibinfo{year}{2021}).
\newblock \urlprefix\url{https://www.nature.com/articles/s41586-021-03588-y}.

\bibitem{fowler_surface_2012}
\bibinfo{author}{Fowler, A.~G.}, \bibinfo{author}{Mariantoni, M.}, \bibinfo{author}{Martinis, J.~M.} \& \bibinfo{author}{Cleland, A.~N.}
\newblock \bibinfo{title}{Surface codes: {Towards} practical large-scale quantum computation}.
\newblock \emph{\bibinfo{journal}{Physical Review A}} \textbf{\bibinfo{volume}{86}}, \bibinfo{pages}{032324} (\bibinfo{year}{2012}).
\newblock \urlprefix\url{https://link.aps.org/doi/10.1103/PhysRevA.86.032324}.

\bibitem{gidney_fault-tolerant_2021}
\bibinfo{author}{Gidney, C.}, \bibinfo{author}{Newman, M.}, \bibinfo{author}{Fowler, A.} \& \bibinfo{author}{Broughton, M.}
\newblock \bibinfo{title}{A {Fault}-{Tolerant} {Honeycomb} {Memory}}.
\newblock \emph{\bibinfo{journal}{Quantum}} \textbf{\bibinfo{volume}{5}}, \bibinfo{pages}{605} (\bibinfo{year}{2021}).
\newblock \urlprefix\url{http://arxiv.org/abs/2108.10457}.
\newblock \bibinfo{note}{ArXiv:2108.10457 [quant-ph]}.

\bibitem{benito_comparative_2024}
\bibinfo{author}{Benito, C.}, \bibinfo{author}{López, E.}, \bibinfo{author}{Peropadre, B.} \& \bibinfo{author}{Bermudez, A.}
\newblock \bibinfo{title}{Comparative study of quantum error correction strategies for the heavy-hexagonal lattice} (\bibinfo{year}{2024}).
\newblock \urlprefix\url{http://arxiv.org/abs/2402.02185}.
\newblock \bibinfo{note}{ArXiv:2402.02185 [quant-ph]}.

\bibitem{acharya_quantum_2024}
\bibinfo{author}{Acharya, R.} \emph{et~al.}
\newblock \bibinfo{title}{Quantum error correction below the surface code threshold} (\bibinfo{year}{2024}).
\newblock \urlprefix\url{http://arxiv.org/abs/2408.13687}.
\newblock \bibinfo{note}{ArXiv:2408.13687 [quant-ph]}.

\end{thebibliography}

\twocolumngrid
\beginsupplement

\clearpage
\onecolumngrid
\begin{center}
	\textbf{\large Supplementary information for\\ 
``Magic State Injection on IBM Quantum Processors Above the Distillation Threshold"}
\end{center}
\bigskip
\twocolumngrid

\section{Asymmetric Feature of Threshold} \label{threshold_sec}
Fig.\,\ref{fig2}a and b in the main text show logical error rates of Z and X errors, respectively. There is a crossover point for each logical error type, known as the threshold value ($p_{\text{th}}$), where the relation between the physical error rate and the corresponding logical error rate changes: For $p < p_{\text{th}}$, $p_L$ decreases as the code distance increases, whereas for $p > p_{\text{th}}$, $p_L$ increases with the code distance. The threshold values for logical phase- ($p_{\text{th}}^Z\approx 0.31\%$) and bit-flip ($p_{\text{th}}^X\approx 0.37\%$) errors are distinct.

\begin{equation}
p^Z_L = C_Z/\Lambda_Z^{a_Z(d+1)/2}, \quad p^X_L = C_X/\Lambda_X^{a_X(d+1)/2}     
\label{threshold}
\end{equation}

To analyze in more detail, we use the fitting equations \eqref{threshold} describing the logical error rate of the Z and X errors as $p^Z_L(p, d)$ and $p^X_L(p, d)$ \cite{google_quantum_ai_exponential_2021}. The fitting is carried out using predefined threshold values for phase- and bit-flip errors denoted as $p^Z_{\text{th}}$ and $p^X_{\text{th}}$, respectively, where $\Lambda_Z \propto p^Z_{\text{th}}/p$ and $\Lambda_Z \propto p^X_{\text{th}}/p$. In these formulas, the terms $C_Z$, $C_X$, $a_Z$, and $a_X$ are fitting constants. Especially, $a_Z$ and $a_X$ are parameters closely relevant to the code performance, specifically associated with the number of errors that form a non-trivial error chain. Even in the regime where the qubit error rate is lower than a threshold, which requires exponential growth in the number of samples, these statistical arguments allow estimating logical error rates using fitting values \cite{fowler_surface_2012,google_quantum_ai_exponential_2021,gidney_fault-tolerant_2021,benito_comparative_2024,acharya_quantum_2024}.

When the parameters ($a_Z$ and $a_X$) are set to 1, the minimum number of errors required to produce a logical error is ${d+1}\over{2}$ for an odd code distance $d$. This shows how the code can efficiently correct bit-flip and phase-flip errors. However, if these parameters are less than 1, it indicates that a logical error could occur with fewer errors. For example, with a code distance of $d=3$, at least two errors are required to cause a logical error. However, if one of the parameters is less than 1, even a single error could induce a logical error of that type, indicating the possibility of a weight-one error transforming into a weight-two error.

\begin{align}
a_Z \approx 0.7, \quad a_X \approx 1 \nonumber
\end{align}

We obtain the fitting parameters as $a_Z \approx 0.7$ and $a_X \approx 1$. Although the code to correct for logical X errors maintains the effectiveness of code distance, it is biased and vulnerable to logical Z errors. We note the asymmetric property in correcting for the two types of errors, even under the unbiased noise model. We attributed this property to the propagation of errors among data qubits.

When layers of CNOT gates are implemented in a sub-round syndrome extraction circuit, errors on data qubits can spread through the gates. Crucially, this error propagation occurs, propagating an error from one data qubit to its vertically adjacent neighbor rather than a horizontally neighboring data qubit. The propagation of errors in the horizontal direction is restricted because of the existence of either syndrome qubits or the constraints of direct interaction. In contrast, errors can propagate vertically due to the process for (un)folding stabilizers requiring long-range CNOT gates among vertically neighboring data qubits. This directional limitation on error propagation affects the efficiency of code-correcting errors, as a weight-one error can escalate into a weight-two error. It is important to note that the detrimental impact is particularly significant for a logical error that coincides with the direction of the data qubits associated with the logical operator, weakening the code's error correction capability in such cases. In our case, the propagation of vertical error hampers the correction of a vertically defined logical error, which is the logical Z operator. Therefore, error propagation among data qubits causes ambiguity in calculating a correction operator and a lower error threshold, decreasing the effective code distance.
\section{Expectation values} \label{expectation_sec}

\begin{figure*}[t]
    \centering
    \includegraphics[width=\textwidth]{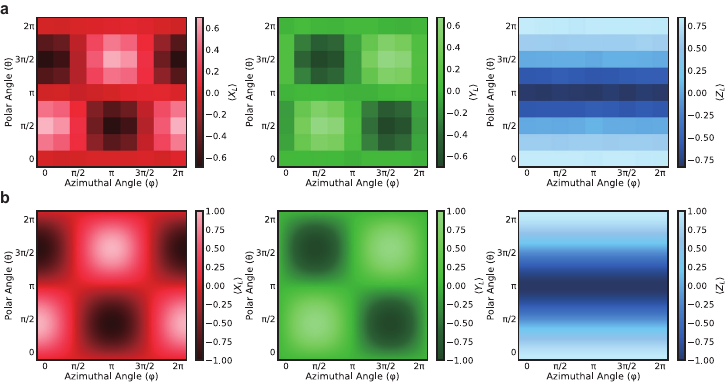}
    \caption{ \textbf{Expectation values.} \textbf{a.} The expectation values of logical Pauli operators X, Y, and Z in raw logical magic states prepared on \texttt{ibm\_fez} device. These values are plotted on a plane with the polar ($\theta$) and azimuthal$ (\phi$) angles, ranging from 0 to $2\pi$. \textbf{b.} Their theoretical values are plotted as functions of parameters $\theta$ and $\phi$ as expressed in \eqref{expectation_X}, \eqref{expectation_Y}, and \eqref{expectation_Z}.}
    \label{fig_app1}
\end{figure*}

Fig.\,\ref{fig_app1}a shows experimentally estimated logical expectation values of logical Pauli operators. The values are estimated from the samples that produce the trivial syndrome obtained by sampling $2 \times 10^4$ times for logical measurements on each Pauli basis. Their ideal values can be expressed in terms of polar($\theta$) and azimuthal($\phi$) angles as shown in Fig.\,\ref{fig_app1}b: 

\begin{align}
\langle\psi_L|\hat{X}_L|\psi_L\rangle &= \text{sin}\theta \text{cos}\phi \label{expectation_X}\\ 
\langle\psi_L|\hat{Y}_L|\psi_L\rangle &= \text{sin}\theta \text{sin}\phi \label{expectation_Y}\\
\langle\psi_L|\hat{Z}_L|\psi_L\rangle &= \text{cos}\theta 
\label{expectation_Z}
\end{align}

Although the experimental results range approximately from $-0.6$ to $0.6$ for the logical X and Y operators and around $-0.8$ to $0.8$ for the logical Z operator, their theoretical values vary from $-1$ to $1$. However, we find that the experimental results closely align with the ideal values.

\begin{figure}[t]
    \centering
    \includegraphics[width=0.5\textwidth]{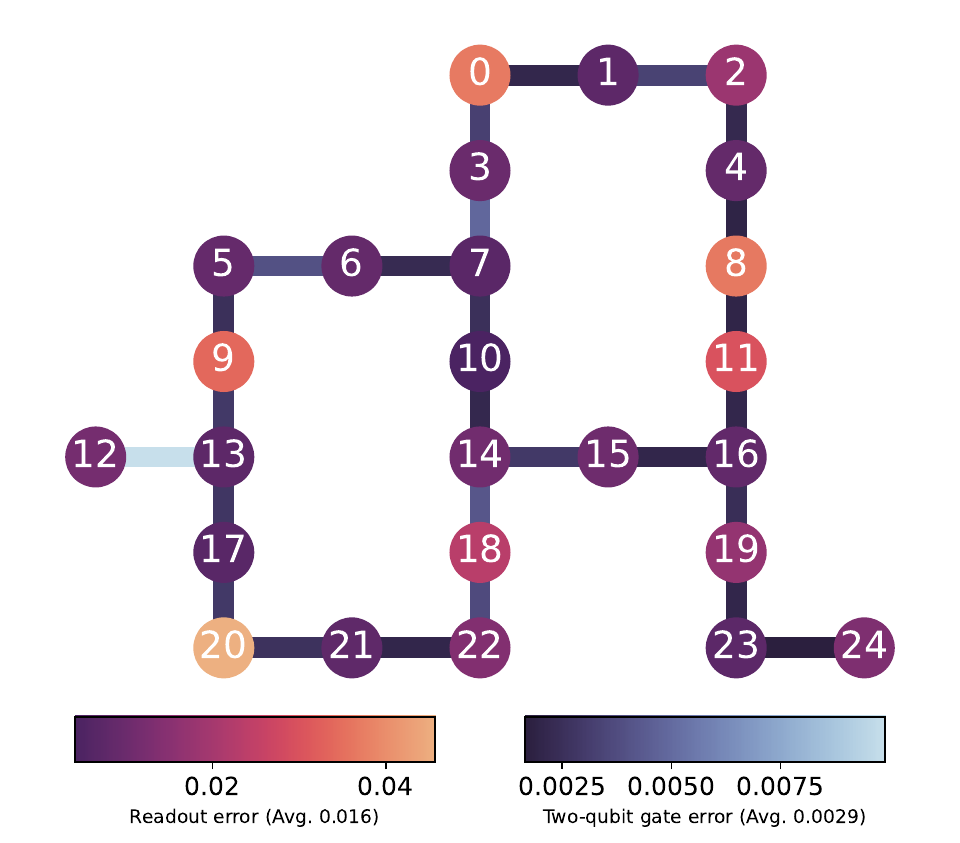}
    \caption{ \textbf{Hardware specifications.} The graph shows the chosen physical qubits’ specifications of the distance-3 rotated surface code onto the heavy-hexagon structure ($\texttt{ibm\_fez}$). Each node and edge correspond to the physical qubit and the connectivity of a two-qubit gate (CZ). The error rates of readout and two-qubit gate are displayed with colors. Their average error rate are $1.6 \times 10^{-2}$ and $2.9 \times 10^{-3}$. }
    \label{fig_app2}
\end{figure} 

\end{document}